\documentclass[rmp,aps,nofootinbib,twocolumn]{revtex4}

\usepackage{graphics}
\usepackage{epsfig}
\usepackage{natbib}
\usepackage{amssymb}
\usepackage{bm,amsmath}

\begin{document}
\title{Colloquium: Physics of optical lattice clocks}

\author{Andrei Derevianko}
\email{andrei@unr.edu}
\affiliation{Department of Physics, University of Nevada, Reno,
Nevada 89557, USA}

\author{Hidetoshi Katori}
\email{katori@amo.t.u-tokyo.ac.jp}
\affiliation{Department of Applied Physics, Graduate School of Engineering, The University of Tokyo,
Bunkyo-ku, 113-8656 Tokyo, Japan}
\affiliation{CREST, Japan Science and Technology Agency, 4-1-8 Honcho Kawaguchi, Saitama, Japan}

\begin{abstract}
Recently invented and demonstrated optical lattice clocks hold  great promise for improving the precision of modern timekeeping.
These clocks aim at the $10^{-18}$ fractional accuracy, which translates
into a clock that would neither lose or gain a fraction of a second over an estimated age of the Universe.
In these clocks, millions of atoms are trapped and interrogated simultaneously, dramatically improving clock stability.
Here we discuss the principles of operation of these clocks and, in particular,  a novel concept of ``magic'' trapping of atoms in optical lattices. We also highlight recently proposed microwave lattice clocks and several applications that employ the optical lattice clocks as a platform for precision measurements and quantum information processing.
\end{abstract}


\maketitle

\section{INTRODUCTION}

Precision timepieces are marvels of human ingenuity.
The earliest surviving clocks, sundials and water clocks, are traced to ancient Egypt~\cite{Ush29}.
The first mechanical clocks were built around the 14th century. The 20th century, with
the advent of quantum mechanics, saw the invention of atomic clocks.
Each qualitative shift in clockwork technology  was accompanied by a dramatic
improvement in timekeeping accuracy.
For example, the water clocks had an error exceeding 15 minutes a day~\cite{Ush29}, while inexpensive quartz crystal clocks may drift a millisecond or so in several days.
The best atomic clock to date~\cite{ChoHumKoe10} may be off by only a fraction of a picosecond a day.
Here we review a novel and rapidly developing class of atomic clocks, optical lattice clocks, which hold a promise of improving accuracy of  modern timekeeping by an order of magnitude.  This translates into an astonishingly accurate clock that would neither lose or gain a fraction of a second over an estimated age of the Universe. In other words, if somebody were to build
such a clock at the Big Bang and if such a timepiece were to survive the 14 billion years, the clock would be off by no more than a mere second. Moreover, compared to other, competing, atomic clocks, the optical lattice clocks promise to reach this accuracy within seconds of integration time.

Over the past half-a-century, the precision time-keeping has been carried out with atomic clocks.
In particular, since 1967, the SI unit of time, the second, is defined as a duration of a certain number of periods of radiation corresponding to the transition between
two hyperfine levels of the ground state of the $^{133}$Cs atom~\cite{NISTSI01}.
Atomic clocks are essential elements of the Global Positioning System (GPS) and
are important for synchronizing signals in digital networks. Fundamental research
ranges from testing the effects of special and general relativity~\cite{Tur09_Q2CBook} to  probing time-variation of fundamental constants~\cite{RosHumSch08}.

 Atomic clocks operate by locking the frequency of an external (e.g., microwave or laser) source in resonance with an internal atomic transition. Counting the number of oscillations at the source tells time. In practice, realizing this scheme requires that the natural frequency of the atomic transition, $\nu_0$, is impervious to external perturbations. Also, there is a certain width of the resonance $\delta \nu$ which limits the uncertainty with which $\nu_0$ may be found. The width of the resonance is
determined, for example, by the inverse of the observation time or ultimately by the natural radiative lifetime of the transition. The relevant parameter characterizing the atomic oscillator is the quality factor (Q-factor), $Q = \nu_0 /\delta \nu$.  Finding the frequency $\nu_0$ precisely requires multiple measurements
on a quantum system.
The relevant indicator of the clock performance is the fractional
instability given by the  Allan deviation~\cite{Ala66}
\begin{equation}
 \sigma_y \approx \frac{1}{Q} \, \frac{1}{\sqrt{N_\mathrm{at} \tau}} \, , \label{Eq:Allan}
\end{equation}
where $N_\mathrm{at}$ is the number of atoms interrogated per unit time and $\tau$ is the total measurement time~\cite{ItaBerBol93}. The stability tells how fast the average of multiple measurements over time approaches the central value.  Good clocks are required to have both excellent accuracy and stability. Therefore, besides  being insensitive to external perturbations, having larger Q-factors, longer  measurement times, and larger atomic samples is beneficial.

All other factors being equal, working with higher frequencies improves the fractional accuracy and the
stability of the clock. This leads to a broad division of  modern atomic clocks into microwave and optical clocks. For example, the $^{133}$Cs standard operates at  $9.2 \times 10^9$ Hz, while the optical Sr lattice clock runs at $4.3 \times 10^{14}$ Hz. The two frequencies differ by 4-5 orders of magnitude.
While the unit of time is presently defined in terms of the microwave Cs standard, the optical clocks
have already outperformed the Cs clocks~\cite{ChoHumKoe10,LudZelCam08}.

Historically, the higher frequency of optical clocks posed a difficulty, as electronic  cycle
counters were not able to cope with optical frequencies. This difficulty was  resolved with the
invention of optical frequency combs, which act as ``optical gears'' and link the optical
clocks to electronic counters~\cite{JonDidRan00,UdeReiHol99}.  Considering this remarkable progress, it is anticipated that eventually the second will be redefined based on the output of optical clocks.

Moving to higher frequencies is advantageous since a number of systematic corrections do not scale
with frequency at all, so there is an immediate improvement in the fractional accuracy. The Doppler shift,
$\delta \nu_\mathrm{D} = - (v/c) \, \nu_0$,  however,
is proportional to the clock frequency; apparently, one needs to reduce atomic velocities $v$ ($c$ is the speed of light),
or, ultimately, trap the atoms. Reducing the velocities additionally increases interrogation time thereby improving the Fourier-limited spectral resolution.

Presently, we may distinguish between two types of competing optical clocks working with trapped species: ion clocks and optical lattice clocks.
In ion clocks
an ion is cooled down to the zero-point energy of the trapping potential. The disadvantage of these clocks is that only a single ion (or only a few ions)
can be used, since trapping multiple ions simultaneously introduces large perturbations of the clock frequency (due to the Coulomb ion-ion interactions ions are pushed out of the trap center where electric field is zero.)
By contrast, the optical lattice clocks (the subject of this Colloquium) employ neutral atoms; the atoms are trapped in specially-engineered standing-wave laser fields termed optical lattices.
Since the interactions between neutral atoms are fairly short-ranged, millions of atoms can be trapped and interrogated simultaneously. This greatly improves the stability of the clock. Qualitatively,
lattice clockwork is equivalent to millions of ion clocks working in parallel.

Trapping atoms with lasers, however, brings a seemingly insurmountable challenge: Optical fields strongly perturb atomic energy levels via the dynamic (or AC) Stark effect - clock frequencies are shifted away from their unperturbed values.
For example, a typical differential Stark  shift induced by trapping a 10-$\mu$K-cold Sr atom exceeds 100 kHz; this translates into a fractional clock accuracy of $10^{-9}$ or so,  which is many orders of magnitude worse than that of the existing clocks.
In addition,
the  Stark shift is proportional to the local intensity of the trapping lasers; the shift is non-uniform across the
atomic ensemble and it is also sensitive to laser intensity fluctuations. So trapping seems to be both advantageous and detrimental for precision optical timekeeping.
%
This dilemma was elegantly resolved using so-called ``magic'' traps~\cite{KatIdoKuw99,McKBucBoo03,YeKimKat08}. At the
``magic'' trapping conditions two levels of interest are shifted by exactly same amount by the trapping fields; therefore the differential effect of trapping fields simply vanishes for the clock transition.

\begin{figure}
\begin{center}
\includegraphics*[width=0.8\linewidth]{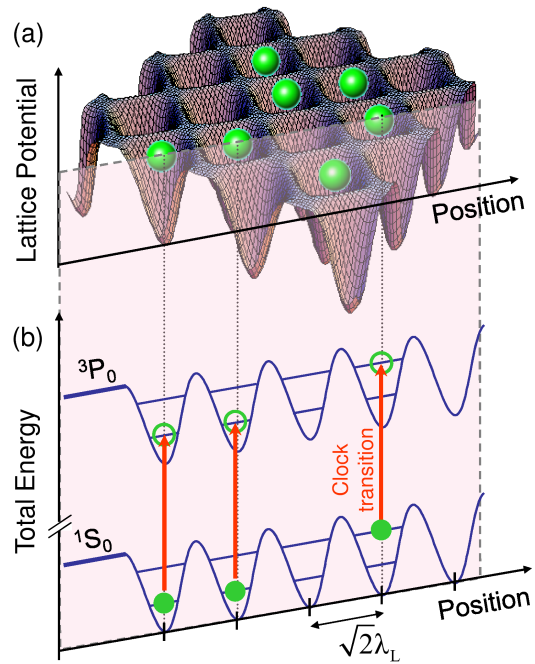}
\end{center}
\caption{(Color online) Illustration of the essential elements of optical lattice clocks. (a) A spatial interference of laser beams creates an egg-carton-like optical potential that traps clock atoms. The atoms are confined to
regions much smaller than the laser wavelength $\lambda_{\rm L}$. (b) Atoms are probed on the $^1\!S_0-{}^3\!P_0$ clock transition. The wavelength $\lambda_{\rm L}$ is tuned to its ``magic'' value so that the clock $^1\!S_0$ and $^3\!P_0$ states are equally energy-shifted by the lattice potential, leaving the transition frequency unperturbed.}
\label{Fig:LatticeClock}
\end{figure}

The optical lattice clocks using the $^1S_0-^3P_0$ transition in alkaline-earth atoms were
proposed in 2002 by Katori~\cite{Kat02}. Figure~\ref{Fig:LatticeClock} illustrates the concept of the clock.
This idea was followed by rapid progress in developing the lattice clocks. A detailed theoretical proposal~\cite{KatTakPal03} for the Sr clock appeared in 2003, the magic wavelength determined experimentally \cite{TakKat03}, and finally the Sr clock was demonstrated just a couple of years later in three different laboratories in Tokyo, Boulder, and Paris~\cite{TakHonHig05,TakHonHig06,LudBoyZel06,LeTBaiFou06}. Recognizing this success, as early as 2006, the Sr optical lattice clock was adopted by the International
Committee for Weights and Measures (CIPM) as one of the secondary representations of the second.
This formalized the Sr clock as a promising candidate for the future redefinition of the second.

The contender status of the lattice clocks was further solidified in 2008 when the international frequency comparison carried out in  Boulder~\cite{CamLudBla08}, Paris~\cite{BaiFouLeT08}, and Tokyo~\cite{HonMusTak09}, agreed with a fractional uncertainty of $6\times 10^{-16}$ that was only limited by the uncertainty of the Cs primary frequency standard.
Similar efforts were undertaken for Yb and Hg lattice clocks. A theoretical analysis of Yb clock performance was carried
out in 2004~\cite{PorDerFor04}; the Yb clock was demonstrated in 2006~\cite{BarHoyOat06}
 and the clock frequency was measured with the accuracy near that of the Cs standard in 2009~\cite{KohYasHos09,LemLudBar09}.
 Hg clock was proposed in 2008~\cite{HacMiyPor08}. The projected fractional accuracy of Hg clocks is at the level of $10^{-18}$ and efforts on building the Hg clock are underway in several laboratories around the world~\cite{PetChiDaw08}.

The fruitful ideas of the optical-frequency-domain lattice clocks were extended to  {\em microwave} frequencies (``microMagic'' clocks). The original proposal~\cite{BelDerDzu08Clock} deals with microwave transition in Al and Ga atoms. For metrologically important Cs and Rb atoms, finding ``magic'' trapping conditions  has proven to be a challenge: an additional control of laser polarizations, magnetic fields and trapping geometry is required~\cite{FlaDzuDer08,LunSchPor10,Der10Bmagic,Der10DoublyMagic}. At the same time, there are compelling benefits to exploring ``microMagic'' clocks. In comparison with the state of the art microwave clocks, the fountain clocks~\cite{WynWey05}, one of the benefits is a much smaller (micrometer-scale) size of volume occupied by clock atoms. Also presently the stability of the primary Cs frequency standard is limited~\cite{SanLauLem99} by the quantum projection noise limit~\cite{ItaBerBol93}, described by Eq.~(\ref{Eq:Allan}).
The stability can be substantially improved by using highly-entangled
ensembles of clock atoms~\cite{LeiBarSch04}; optical-lattice-based ``microMagic'' clocks are excellent candidates for realizing such ideas.


This colloquium is organized as follows. In Section~\ref{Sec:StarkShifts}, we review the interaction of atoms with  off-resonant laser light, describe determination of magic wavelengths and spectroscopy in optical lattices. In Section~\ref{Sec:Optical}, we discuss operation of optical lattice clocks and their error budget. In Section~\ref{Sec:MuMagic} we discuss a recently proposed class of atomic {\em microwave} clocks (``microMagic'' clocks). Finally, in Section~\ref{Sec:Beyond}, we highlight several proposals for using  lattice clocks in precision measurements
and quantum information processing.

\section{OPTICAL LATTICES AND MAGIC WAVELENGTH}
\label{Sec:StarkShifts}
In this section, we provide the introductory background required for understanding the basic physics of optical lattice
clocks: optical lattices, ``magic'' wavelengths, and the Lamb-Dicke spectroscopy.

\subsection{Light shifts and polarizabilities}
The key idea for realizing the lattice clock is the concept of ``magic'' trapping. Generally,
magic optical trapping potentials for a specific clock transition may be defined as specially tailored trapping fields in which differential shift of the clock transition vanishes exactly. Notice that the individual levels may be perturbed by the trapping fields very strongly. Nevertheless, at the magic conditions both clock levels are shifted identically.

The effect of optical (laser) trapping fields on a given level is quantified using the AC Stark shift and dynamic polarizability. The {\em  static} Stark shift is a familiar concept in quantum mechanics; it refers to a shift of energy levels in the presence of externally-applied static electric fields. For states of definite parity, the leading contribution arises in the second order of perturbation theory, and it is quadratic in the E-field. The coefficient of proportionality is called the static dipole polarizability. When the electric field oscillates, like in lasers, the Stark shift of energy levels remains time-independent. This is similar to the Lamb shift, where time-dependent vacuum fluctuations lead to a static shift. An interested reader is referred to a review~\cite{ManOvsRap86} for details.
The theoretical analysis is similar to the static E-field case; in particular the static polarizability is replaced by the {\em dynamic} polarizability.

An order-by-order expansion of the AC Stark (light) shift of the energy of level $a$ reads
\begin{equation}
 \delta E^\mathrm{Stark}_a = - \alpha_a(\omega_L) \left( \frac{\mathcal{E}_L}{2} \right) ^2
 - \alpha'_a(\omega_L) \left( \frac{\mathcal{E}_L}{2} \right) ^4 +  O(\mathcal{E}_L^6) \,,
 \label{Eq:StarkShift}
\end{equation}
where $\mathcal{E}_L$ and $\omega_L$ are the (real-valued) amplitude and the frequency of the laser field. The frequency-dependent quantities
$\alpha_a(\omega_L)$  and $\alpha'_a(\omega_L)$ are the dynamic polarizability and hyperpolarizability, respectively.  The resulting differential Stark shift of the clock frequency is
\begin{equation}
 h \, \Delta \nu^\mathrm{Stark} =  - \Delta \alpha(\omega_L)\, \left( \frac{\mathcal{E}_L}{2} \right) ^2
 - \Delta \alpha'(\omega_L) \left( \frac{\mathcal{E}_L}{2} \right) ^4 +  O(\mathcal{E}_L^6) \,,
  \label{Eq:DiffStarkShift}
\end{equation}
where differential polarizabilities of the two ( the lower $|g\rangle$ and the upper $|e\rangle$) clock levels are
defined as  $\Delta \alpha(\omega_L) = \alpha_e(\omega_L) - \alpha_g(\omega_L)$.
Notice that the hyperpolarizability correction, being of higher order in the electromagnetic coupling,
is relatively small (we will return to this discussion later as the relevant correction has an effect on clock's accuracy). Therefore,  the magic laser  wavelength $\lambda_m$ (or frequency $\omega_m=2 \pi c/\lambda_m$) is determined by computing  dynamic polarizabilities
for the two clock levels as a function of $\omega_L$. Intersections of the two curves determines
the values of $\omega_m$.

The polarizability depends on atomic electric-dipole $D$ matrix elements
and energies $E$ and also on the (generally complex-valued) polarization vector $\hat{\varepsilon}$ of the laser
\begin{equation}
\alpha_a(\omega) =
\sum_{b}\frac{
\left\vert \langle a|\mathbf{D} \cdot \hat{\varepsilon}|b\rangle\right\vert ^{2}}{E_{b}- E_{a}-\omega }+\sum_{b}\frac{
\left\vert \langle a|\mathbf{D} \cdot \hat{\varepsilon}|b\rangle\right\vert ^{2}}
{E_{b}- E_{a} +\omega }.\label{Eq:PolarizabilityGeneral}%
\end{equation}
The sums are over a complete atomic eigen-set.

We may decompose the polarizability for a state $|n FM_F\rangle$ of the total angular momentum $F$ and its projection $M_F$ ($n$ encompasses all remaining quantum numbers) into the following contributions,
\begin{eqnarray}
\lefteqn{\alpha_{nFM_F}(\omega)=  \alpha_{F}^{S} (\omega) +(\hat{k}\cdot\hat{B}) \mathcal{A}\frac{M_{F}}{2F}\alpha_{F}^{V} (\omega) +} \nonumber \\
&& \frac{1}{2}\,\left(3\left|\hat{\varepsilon}\cdot\hat{B}\right|^{2}-1\right)
        \frac{3{M_{F}}^{2}-F(F+1)}{F(2F-1)}\alpha_{F}^{T} (\omega)  \, .
\label{Eq:alphaBreakDown}
\end{eqnarray}
Here the superscripts $S$, $V$, and $T$ distinguish the scalar, vector, and tensor parts of the polarizability.  $\hat{k}$ and $\hat{B}$ are the unit vectors along the lattice wavevector and quantizing B-field, respectively.  $\mathcal{A}$ is the degree of circular polarization of the light: $\mathcal{A} = \pm 1$ for $\sigma_\pm$ light.  For a linearly polarized laser  $\mathcal{A} = 0$  and the vector contribution drops out. $\hat{\varepsilon}$ is complex for non-zero degree of circular polarization.

Arriving at the tensorial decomposition (\ref{Eq:alphaBreakDown}) of the  polarizability (\ref{Eq:PolarizabilityGeneral}) requires techniques of quantum theory of angular momentum~\cite{VarMosKhe88}. Qualitatively, the decomposition  could be understood from the following arguments. Eq.(\ref{Eq:PolarizabilityGeneral}) contains four vectors: two polarizations $\hat{\varepsilon}$ and two dipole moments $\mathbf{D}$
in a particular (rotationally-invariant) combination of cartesian components of these vectors:
$\sum_{ij} \varepsilon_i D_i \mathcal{R}\varepsilon_j^*D_j=\sum_{ij} (\varepsilon_i \varepsilon_j^*)  (D_i \mathcal{R} D_j) \equiv \sum_{ij} \mathcal{P}_{ij} \mathcal{D}_{ij}$, where $\mathcal{R}=\sum_b
|b\rangle \langle b | (E_{b}- E_{a} \pm \omega)^{-1}$
is the resolvent operator. Let's focus on the combinations $\mathcal{D}_{ij}=D_i \mathcal{R} D_j$. These form components of a rank-2 cartesian polarizability tensor $\mathcal{D}_{ij}$. In general, it may be decomposed into three irreducible parts: a scalar $\mathcal{D}^{(0)}=1/3(\mathbf{D} \cdot \mathcal{R} \mathbf{D})$, a vector $\mathcal{D}^{(1)}=1/2 \, \mathbf{D}\times \mathcal{R}\mathbf{D}$,
and a symmetric traceless tensor $\mathcal{D}^{(2)}_{ij}=1/2 \left( \mathcal{D}_{ij} +  \mathcal{D}_{ji} \right)-\delta_{ij} \mathcal{D}^{(0)}$. Similar decomposition may be carried out for the polarization
tensor $\mathcal{P}_{ij} =\varepsilon_i \varepsilon_j^*$. Combining irreducible components of the polarizability and polarization tensors, we arrive at the three (scalar, vector, and tensor) contributions to Eq.~(\ref{Eq:alphaBreakDown}).

Further, evaluation of matrix elements of irreducible polarizability tensors $\mathcal{D}^{(L)}$ in atomic basis $|n FM_F\rangle$  is aided by the Wigner-Eckart theorem, which states that
a matrix element may be factorized into two parts,
dependent and independent on magnetic quantum numbers $M_F$ of atomic states.
The former gives rise to $M_F$-dependent prefactors in vector and tensor contributions to Eq.~(\ref{Eq:alphaBreakDown}) and the latter is encapsulated
in $M_F$-independent (or ``reduced'') polarizabilities $\alpha_{F}^{S} (\omega)$, $\alpha_{F}^{V} (\omega)$, $\alpha_{F}^{T} (\omega)$. The $M_F$-dependent prefactors from the Wigner-Eckar theorem fix angular selection rules for matrix elements (these are suppressed in Eq.(\ref{Eq:alphaBreakDown})): the total angular momentum $F$ is to be greater or equal to $L/2$ for non-vanishing diagonal matrix element of  tensor $\mathcal{D}^{(L)}$. Finally, specifying $M_F$ fixes direction of
quantization axis and the angular factors in Eq.~(\ref{Eq:alphaBreakDown}) arise
when evaluating irreducible polarization tensors in this fixed reference frame.

\subsection{Theoretical determination of magic wavelengths}
\begin{figure}[h]
\begin{center}
\includegraphics*[width=0.8\linewidth]{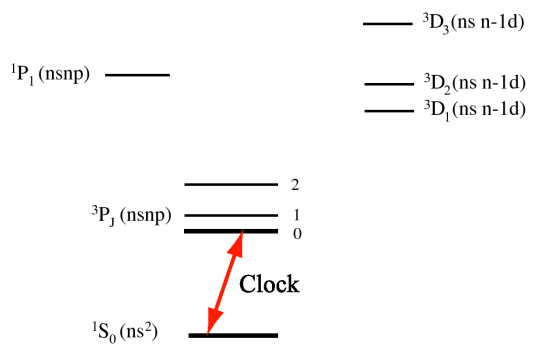}
\end{center}
\caption{ (Color online) A diagram of the low-lying energy levels  for  Mg ($n$=3),
Ca ($n$=4), Sr ($n$=5), and Yb ($n$=6). The relative position of the
levels above the $^3P_J$ fine-structure manifold depends on the
atom. This diagram reflects energy levels of Yb (core-excited
states are not shown). The clock transition is between the ground
and the lowest-energy $^3P_0$ state. }
\label{Fig:Levels}
\end{figure}

Certainly, the values of magic wavelength depend on specific atoms. The clockwork in optical lattice clocks  takes advantage
of the electronic structure of atoms with two valence electrons
outside a closed-shell core. Such systems include group II and IIb atoms,
such as magnesium, calcium, and strontium, or more complex divalent
atoms such as ytterbium and mercury atoms. A typical level structure
of such atoms is shown in Fig.~\ref{Fig:Levels}. The clock
transition is between the ground  $ns^2\,^1\!S_0$ state and the
$J=0$ component of the lowest-energy triplet state fine-structure
manifold, $nsnp\,^3\!P_J$. The scalar character of the $J=0$ clock states
makes the clock transition insensitive to magnetic fields and vector light shift perturbations (see below).

\begin{figure}[h]
\begin{center}
\includegraphics*[width=0.8\linewidth]{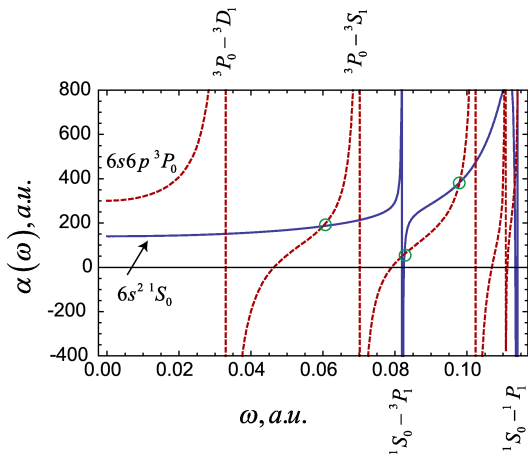}
\caption{(Color online) Dynamic polarizabilities $\alpha$ of the two clock levels in Yb as a function of laser frequency $\omega$. Blue solid curve is the polarizability of the $6s^2\,^3\!P_0$ lower clock state and red dashed line is $\alpha(\omega)$ of the $6s6p\,^3\!P_0$ upper clock state.
The a.u.\ stands for atomic units. Conversion factors are:
 $\alpha/h [\mathrm{Hz/(V/m)}^2] = 2.48832 \times  10^{-8} \alpha [\mathrm{a.u}]$ for polarizability
and $ \omega/(2\pi) [\mathrm{Hz}] = 4.1341 \times 10^{16} \omega [\mathrm{a.u.}]$ for frequency.
``Magic'' frequencies of the laser field are marked by small circles on the plot.
}
\label{Fig:YbMagic}
\end{center}
\end{figure}

Evaluation of  polarizabilities involves summing over electric-dipole-allowed transitions:
for example, in Fig.~\ref{Fig:Levels}, the sum for the ground state  includes the $^3P_1$, $^1P_1$ and higher-energy $J=1$ odd-parity states (not shown). The upper clock level has the parity opposite to that of the ground state and the intermediate states will include $J=1$ even-parity state, such as the $^3D_1$ state in Fig.~\ref{Fig:Levels}. In Fig.~\ref{Fig:YbMagic}, we present results of such calculations~\cite{DzuDer10} for Yb atom. The two polarizabilities of the clock states spike at resonances. The resonant transitions are marked on the plot. At the lower-end frequency range, $\omega < 0.08 \, \mathrm{a.u.}$, the polarizability of the $^3P_0$ state goes through two resonances,
while  $\alpha_{^1\!S_0}$ remains relatively flat. This dissimilar behavior of the two polarizabilities almost inevitably results in crossings of the two curves: values of laser wavelengths at these crossings are ``magic''.

There are several magic wavelengths predicted from crossings of polarizabilities in
Fig.~\ref{Fig:YbMagic}; the first five $\lambda_m$ are tabulated in~\cite{DzuDer10}.
The experimentally realized Yb clock operates at the first and longest wavelength
$\lambda_m \approx 759 \, \mathrm{nm}$. This wavelength was theoretically predicted in~\cite{PorDerFor04} and subsequently measured in~\cite{BarHoyOat06,BarStaLem08}.

\subsection{Optical lattices and Lamb-Dicke spectroscopy}
\label{Sec:OptLat}
Before discussing experimental determination of magic wavelengths, we would like to
introduce several basic ideas of trapping and spectroscopy in optical lattices.
Let us consider two counter-propagating laser beams of linear
polarization and of the same wavelength $\lambda_L$ and intensity $I_L=c/(8\pi) \mathcal{E}_L^2$. The resulting standing wave has
the intensity nodes separated by $\lambda_L/2$.
This oscillatory intensity pattern translates into spatially modulated
Stark shift of the energy levels via Eq.~(\ref{Eq:StarkShift}) or, equivalently, to the optical potential experienced by the atom
\begin{equation}
U\left(  r,z\right)   = U_0 \,
\exp\left\{  -2 \left(r/w(z)\right)^{2}\right\}
\cos^{2}\left(  2\pi z/\lambda_{L}\right) \, . \label{Eq:Uopt1D}
\end{equation}
Here the z-axis lies along the laser beam, $r$ is the radial coordinate in the
transverse direction and $w(z)$ is the beam waist. This geometry is conventionally referred to as the 1D lattice.
The potential depth, $U_0$,
is expressed as
\begin{equation}
U_0   = - \frac{8 \pi}{c} \,\alpha\left(  \omega_{L}\right) I_{L} \, .
\end{equation}
We see that the polarizability $\alpha\left(  \omega_{L}\right)$
governs the  Stark clock shift and also the atomic trapping potential (of course, both the Stark shift and the optical potential describe the very same energy shift). Notice
that since at the magic wavelength the polarizabilities of the two clock states are equal,
both states experience identical trapping potentials.

\begin{figure}
\begin{center}
\includegraphics*[width=0.6\linewidth]{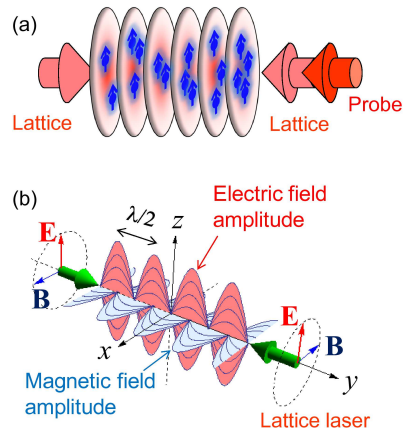}
\end{center}
\caption{(Color online) (a) A one-dimensional optical clock is realized by a standing wave of light tuned to the magic wavelength. Multiply trapped spin-polarized fermions in a single pancake potential may be protected from collisions by the Pauli blocking. (b) Electric and magnetic field amplitudes in a standing wave.}
\label{Fig:1D}
\end{figure}

From Fig.~\ref{Fig:YbMagic}, we see that the polarizability may accept both positive and negative
values. For $\alpha > 0$, $U_0 < 0$ and  atoms are attracted to maxima of local intensity: trapped atoms form layers of pancake-like clouds separated by $\lambda_L/2$ in the axial direction (see Fig.~\ref{Fig:1D}).  By contrast, for the negative values of polarizabilities, the atoms are pushed
to the minima of intensity: they simply escape the 1D lattice along the radial direction.
In this case the confinement can be provided by  3D optical lattices, where three overlapping 1D lattices are oriented along three spatially-orthogonal directions.

The potential (\ref{Eq:Uopt1D}) is periodic in the axial direction. Although the solutions of the corresponding Schrodinger equation for atomic motion can be found in terms of the Wannier and Bloch functions familiar
from solid-state physics, a qualitative consideration will suffice for our goals (see \cite{Lem09} for details). Near the bottom of the wells, the potential is harmonic, with the spacing
between the levels given by
\begin{equation}
\omega_\mathrm{ho}   =\frac{2\pi}{\lambda_{L}} \, \left(  \frac{2 |U_0| }{M} \right)^{1/2} \, ,
\end{equation}
where $M$ is the atomic mass.
Notice that an atom initially trapped in one of the sites may tunnel out to the neighboring wells. As we move up the vibrational energy ladder, the tunneling rate (Bloch bandwidth) increases.
As we interrogate the clock transition of the trapped atoms with a laser of frequency $\omega_p$, the absorbed photon imparts a momentum kick $p=\hbar \omega_p /c$ to the atom. As long as the recoil energy $E_r = p^2/(2M)$ is much smaller than the spacing $\hbar \omega_\mathrm{ho}$ between the harmonic levels, the atoms remain in the same motional state. Thus the absorbed frequency is equal to the internal atomic frequency within the width of the trapped level (we imply trapping in a magic lattice to remove differential Stark shifts). This is the Lamb-Dicke regime\cite{Dic53}, which guaranties that the quantized atomic motion in a trap does not alter the clock frequency.

Classically, a trapped atom oscillates in the optical potential with a frequency $\omega_\mathrm{ho}$. As the atom moves, the electric field of the probe laser experienced by the atom becomes phase modulated: $\mathcal{E}_p = \mathcal{E}_p^0 \cos ({\bf k}_p\cdot {\bf x} \sin \omega_\mathrm{ho}  t - \omega _p t)$ with ${\bf x}$ being the amplitude of atomic oscillation.
When the modulation index $m = {\bf k}_p\cdot {\bf x}\ll  1$, the atom observes an electric field $\mathcal{E}_p \approx \mathcal{E}_p^0 \left\{ {\cos \omega _p t + \frac{m}{2}\left[ {\cos (\omega _p  - \Omega )t - \cos (\omega _p  + \Omega )t} \right]} \right\}$;
 i.e., the field is composed of a central carrier at $\omega_p$ frequency and two weak sidebands.  Clearly, the Doppler shift becomes discretized and is removed in the spectroscopic measurement. The  photon recoil shift is absorbed by the macroscopic objects (lattice) as in the M\"{o}ssbauer effect.


\subsection{Experimental determination of magic wavelength}

\begin{figure}
\begin{center}
\includegraphics*[width=\linewidth]{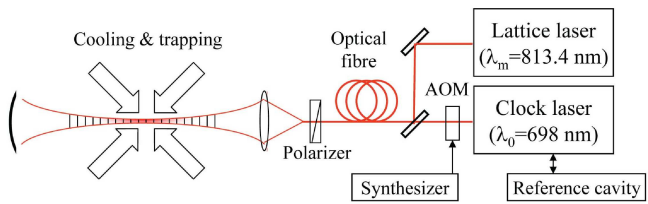}
\end{center}
\caption{(Color online) Schematic diagram of the experimental setup for Sr spectroscopy used in an early experiment~\cite{TakKat03}. Ultracold $^{87}$Sr atoms are loaded into a 1D optical lattice produced by the standing wave of a Ti-sapphire laser tuned to the magic wavelength. 
The atoms interact with the clock laser propagating along this axis and the Lamb-Dicke condition is satisfied. AOM, acousto-optic modulator.}
\label{Fig:SrSetup1}
\end{figure}

\begin{figure}
\begin{center}
\includegraphics*[width=0.7\linewidth]{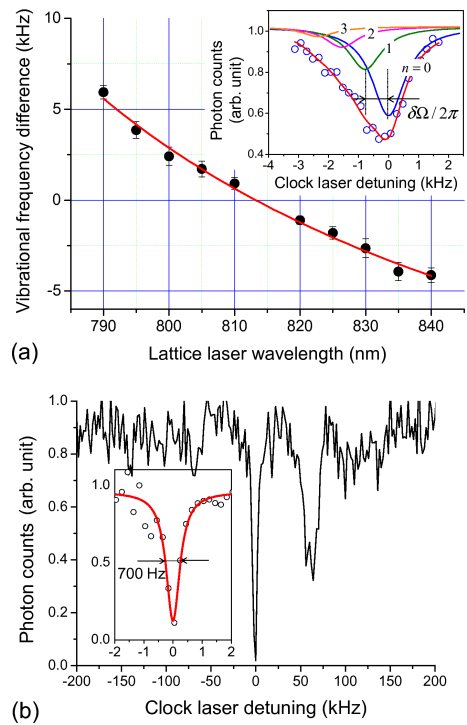}
\end{center}
\caption{(Color online) (a) The $^{87}$Sr magic wavelength was determined by investigating the spectral line broadening of the clock transition, as shown in the inset. This broadening  revealed  the vibrational frequency differences in two states of the clock transition, which is plotted as a function of lattice laser wavelength to determine the degenerate wavelength to be $\lambda_L=813.5 \pm 0.9$ nm.
(b) The first spectrum of the clock transition  in the ``magic lattice''. The spectrum consists of the heating and cooling sidebands at $\approx \pm 64\ {\rm kHz}$ and  the recoilless spectrum (the carrier component) with a linewidth of 700 Hz (FWHM) as shown in the inset.}
\label{Fig:Spectroscopy}
\end{figure}

Figure~\ref{Fig:SrSetup1} shows an experimental setup~\cite{TakKat03} used for clock spectroscopy and determination of magic wavelength. In this experiment,
$^{87}$Sr atoms were laser-cooled and trapped on the $^{1}\!S_0-{}^{3}\!P_1$ transition with a Dynamic Magneto-Optical Trapping  technique \cite{MukKatIdo03}. Roughly 10$^{4}$ atoms with a temperature of about $2~\mu$K were loaded into a 20-$\mu$K-deep 1D optical lattice that was formed by the standing wave of a lattice laser. The atoms were trapped in the Lamb-Dicke regime along the axial direction. The magic wavelength was first determined to be 813.5(9)~nm by investigating the narrowing of the clock spectra as a result of the cancellation of the light shift (see Fig.~\ref{Fig:Spectroscopy}(a)).
At this magic wavelength, the observed  clock spectrum is shown in  Fig.~\ref{Fig:Spectroscopy}(b). The spectrum consists of a central narrow carrier of 700~Hz linewidth (see inset) and two sidebands at $\approx \pm 64$~kHz.
Presently, the magic wavelength for $^{87}$Sr is measured with a 7-digit accuracy~\cite{CamLudBla08}.
It is worth noting that knowing $\lambda_m$ with a mere 7-digit accuracy affects the clock frequency $\nu_0$ only at the 16th significant digit~\cite{KatTakPal03}.
The magic wavelength for $^{171}$Yb was  determined with a similar accuracy~\cite{LemLudBar09}.

\subsection{Higher-order corrections to the Stark shift and blue-detuned lattices}
Our discussion of the magic wavelength trapping focused on the cancellation of the leading, second-order, light shifts in Eq.~(\ref{Eq:DiffStarkShift}). In general, however the fourth and the higher order differential AC Stark shifts cannot be canceled out at the magic wavelength. Their effect on the clock accuracy is of a serious concern.
The fractional shifts of the clock frequency due to hyperpolarizability are predicted to be in the range of $10^{-17}-10^{-19}$ for Sr-, Yb-, and Hg-based optical lattice clocks operating at their magic wavelengths\cite{KatTakPal03,PorDerFor04,HacMiyPor08}.
The effects of hyperpolarizability have been experimentally investigated in Sr~\cite{BruLeTBai06} and Yb~\cite{BarStaLem08} confirming the fractional correction to be less than $10^{-17}$. While at the present level of uncertainty such corrections are affordable, these become important when targeting the $10^{-18}$ uncertainty level.

The detrimental effects of hyperpolarizability can be suppressed by employing  so-called ``blue-detuned'' lattices.
Near an atomic resonance, the polarizability, Eq.~(\ref{Eq:PolarizabilityGeneral}), is dominated by a single contribution.
As seen from Fig.~\ref{Fig:YbMagic}, below (on the red side of) the resonance, the polarizability is positive and above (on the blue side of)  the resonance, $\alpha(\omega)<0$. So far we focused on the red-detuned magic lattices. As discussed in Sec.~\ref{Sec:OptLat}, these trap atoms near the intensity maxima, i.e., at the anti-nodes of the standing wave.
However, when $\alpha_g(\omega_m)=\alpha_e(\omega_m) <0 $, the atoms are confined at the intensity minima of the electric field. Then, for a strong confinement, the intensity averaged over the atomic center-of-mass motion becomes a small fraction of the maximum laser lattice intensity, thereby suppressing the contributions of hyper- and higher-order polarizabilities to the clock shift, Eq.~(\ref{Eq:DiffStarkShift}).

\begin{figure}
\begin{center}
\includegraphics*[width=\linewidth]{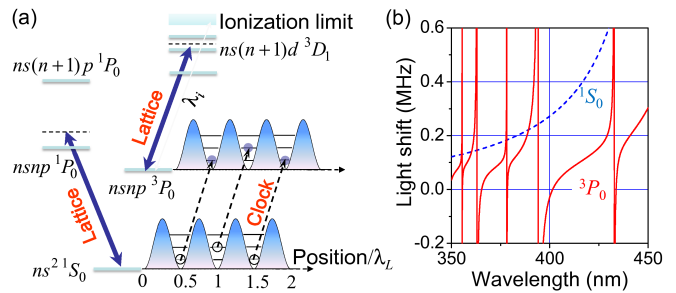}
\end{center}
\caption{(Color online) (a) Energy levels for alkaline-earth atoms relevant to  ``blue-detuned'' magic wavelengths for the $^1S_0-{}^3P_0$  clock transition.
By applying the lattice laser detuned slightly above the nearby resonant
state, $nsnp\,{}^1P_1$ and $ns(n+1)d\,{}^3D_1$, atoms can be trapped
near the nodes of the standing wave. (b) The light shifts for the $^1S_0$ (dashed blue) and
$^3P_0$ (solid red) states of Sr as a function of the lattice laser
wavelength  for a laser intensity of $I=10~{\rm kW/cm}^2$. Intersections of the curves indicate blue magic
wavelengths, which include $\lambda_{\rm b}\approx 360$ and 390 nm.}
\label{blue}
\end{figure}

Determination of the ``blue-detuned'' magic wavelength for Sr is illustrated in Fig.~\ref{blue}.
The desired magic wavelength is located on the blue side of the $5s^2\,^1S_0-5s5p\, ^1P_1$  461 nm transition. A very far-off-resonance condition  is generally difficult to satisfy because the magic wavelength can only be found close to the transition originating from the $5s5p\, ^3P_0$ state, as indicated in Fig.~\ref{blue}(a).
One such wavelength is found at $\lambda_L\approx  390$ nm on the blue side of the $5s5p\, ^3P_0-5s6d\, ^3D_1$ transition at 394 nm. For this magic wavelength, the laser intensity of $I_L = 10$ kW/cm$^2$ yields a trap depth of about 200 kHz, as indicated in Fig.~\ref{blue}(b). The effective light intensity that atoms experience is about one-tenth of the maximum intensity, as the atoms are trapped near the node of the standing wave. With a trap depth of 10 $\mu$K, the 4th-order light shift (the second term in Eq.~(\ref{Eq:DiffStarkShift})) is estimated to be 0.1~mHz, corresponding to a fractional uncertainty of $2\times 10^{-19}$.
The blue magic wavelength for $^{87}$Sr was measured to be 389.889(9) nm \cite{TakKatMar09} by investigating the light shift in a 1D optical lattice  operated at the (red-detuned) magic wavelength of $\lambda_m= 813.4$ nm.

\subsection{Multipolar interactions of atoms with lattice field and atomic-motion-insensitive wavelength}
Our preceding discussion of the Stark shift included only the dominant electric-dipole (E1) interaction with the laser field.
A multipolar expansion of the laser field about the atomic nucleus results in a series of electric (EJ) and magnetic (MJ) multipoles (here $J$ is the tensor rank of the relevant $2^J$-pole operator). The expressions for multipolar polarizabilities are similar to Eq.~(\ref{Eq:PolarizabilityGeneral}) but with
E1 operators replaced by multipolar operators. Although higher-order multipoles are suppressed compared to the E1 contribution, they do affect atomic trapping and magic wavelengths.
 Now let us take into account higher multipoles.
Consider, for example, the linearly polarized ($||{\bf{e}}_z $) standing wave electric field ${\bf{E}} = {\bf{e}}_z E_0 \sin ky\cos \omega t$ with a wave number $k$ and a frequency $\omega$, as shown in Fig.~\ref{Fig:1D}(b).
Following the Maxwell equation $\nabla  \times {\bf{\mathcal{E}}} =  - \frac{1}{c}\frac{{\partial {\bf{\mathcal{B}}}}}{{\partial t}}$ with $c$  the speed of light, the corresponding magnetic field is given by ${\bf{\mathcal{B}}} =  - {\bf{e}}_x \mathcal{E}_0 \cos ky\sin \omega t$. This indicates that the electric and magnetic field amplitudes are one quarter of the wavelength  $\lambda/4=\pi c/(2\omega)$ out of phase in space. Consequently, the magnetic dipole (M1) interaction is maximum at the nodes of the electric field. Furthermore, as the electric quadrupole (E2) interaction is proportional to the gradient of the electric field, the E2 interaction is also maximum at the node of the electric field.
While optical lattice clocks operated at the blue-detuned magic wavelength do minimize the E1 interactions of atoms with the lattice laser field, such lattices are not necessarily free of multipolar light shift perturbations.

The energy shift of atoms in the optical lattices is obtained in the second-order perturbation theory in the E1, M1, and E2 interactions. These vary as $V_{{\rm{E1}}} \sin ^2 ky$, $V_{{\rm{M1}}} \cos ^2 ky$, and $V_{{\rm{E2}}} \cos ^2 ky$.
As a result, it is no longer possible to perfectly match the total light shift in two clock states.
For example, at the magic wavelength for the E1 interaction as discussed previously, differential light shifts due to M1 and E2 interactions exist, which introduce an atomic-motion-dependent light shift because of their spatial mismatch with the E1 interaction \cite{TaiYudOvs08}.



Although the contributions of the M1 and E2 interactions are 6-7 orders of magnitude smaller than that of the E1 interaction in optical lattice clocks \cite{KatTakPal03,PorDerFor04}, they have a non-negligible contribution in pursuing the $1\times 10^{-18}$ level uncertainty;
therefore, a more precise definition of the magic wavelength, including multipolar interactions, is necessary.
Assuming the differential polarizabilities of the E1, M1, and E2 interactions in the clock transition to be  $\Delta \alpha _{{\rm{E1}}} (\lambda_L )$, $\Delta \alpha _{{\rm{M1}}} (\lambda_L )$, and $\Delta \alpha _{{\rm{E2}}} (\lambda_L )$, and the corresponding spatial distributions to be $q_{{\rm{E1}}} ({\bf{r}})$, $q_{{\rm{M1}}} ({\bf{r}})$, and $q_{{\rm{E2}}} ({\bf{r}})$, the transition frequency of atoms in the optical lattices can be given by
\begin{eqnarray}
 \nu (\lambda _L ) = \nu _{\rm{0}}  - \frac{1}{{2h}}[\Delta \alpha _{{\rm{E1}}} (\lambda _L )q_{{\rm{E1}}} ({\bf{r}})
+ \Delta \alpha _{{\rm{M1}}} (\lambda _L )q_{{\rm{M1}}} ({\bf{r}})  \nonumber\\
  + \Delta \alpha _{{\rm{E2}}} (\lambda _L )q_{{\rm{E2}}} ({\bf{r}})]\mathcal{E}^2,
\label{magic1}
\end{eqnarray}
which corresponds to Eq.~(\ref{Eq:DiffStarkShift}), but with the 4th- and higher-order terms omitted.

Atomic-motion-dependent light shift caused by multipolar interactions can be eliminated by choosing particular 3D optical lattice geometries that make the M1 and/or E2 interactions in phase or out of phase with respect to the spatial dependence of E1 interaction \cite{KatHasIli09}.
For example, in the case of a 1D lattice with the E1 spatial dependence  $q_{{\rm{E1}}} ({\bf{r}}) = \sin ^2 ky\left( { = 1 - \cos ^2 ky} \right)$, the corresponding  M1 and E2 interactions may be expressed as  $q_{{\rm{M1}}} ({\bf{r}}) = q_{{\rm{E2}}} ({\bf{r}}) = \cos ^2 ky = \Delta q - q_{{\rm{E1}}} ({\bf{r}})$ with $\Delta q = 1$.
Therefore, by taking $\Delta \alpha _{{\rm{EM}}}  \equiv \Delta \alpha _{{\rm{E1}}}  - \Delta \alpha _{{\rm{M1}}}  - \Delta \alpha _{{\rm{E2}}} $ and $\Delta \alpha _0  \equiv \Delta \alpha _{{\rm{M1}}}  + \Delta \alpha _{{\rm{E2}}} $, Eq.~(\ref{magic1}) can be rewritten as
\begin{equation}
\nu(\lambda _L ) = \nu _{\rm{0}}  - \frac{1}{2h}\Delta \alpha _{{\rm{EM}}} (\lambda _L )q_{{\rm{E1}}} ({\bf{r}})\mathcal{E}^2  - \frac{1}{2h}\Delta \alpha _{\rm{0}} (\lambda _L )\Delta q \mathcal{E}^2,
\label{magic2}
\end{equation}
where the second term on the right-hand side varies in phase with the E1 interaction.
This equation suggests that the precise definition of the magic wavelength is to be an ``atomic-motion insensitive''  wavelength.
The last term provides a spatially constant offset typically at the 10 mHz level and is solely dependent on the total laser intensity $\propto \Delta q \mathcal{E}_0^2 $ used to form the lattice.
This offset frequency can be accurately determined by measuring the atomic vibrational frequencies in the lattice.


%
%
%
%
%

\section{OPTICAL LATTICE CLOCKS}
\label{Sec:Optical}

Armed with the understanding of ``magic-wavelength'' trapping and spectroscopy in the Lamb-Dicke regime,
in this Section we focus on operation of optical lattice clocks and their error budget.

\begin{figure}
\begin{center}
\includegraphics*[width=0.8\linewidth]{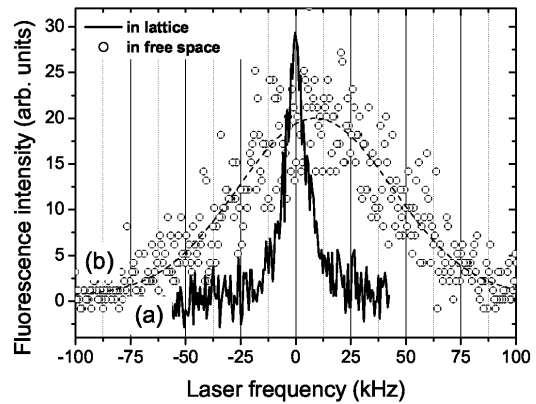}
\end{center}
\caption{Laser-induced fluorescence of atoms (a) confined in a
1D optical lattice and (b) in free fall. The dashed line shows a
Gaussian fit to the data points (b). The confinement suppressed
the Doppler width of 83 kHz and gave a narrow Lorentzian
linewidth of 11 kHz, which was limited by the saturation
broadening. A slight blue shift of the center frequency in (b)
is caused by the photon recoil shift.}
\label{Fig:LDR}
\end{figure}

The first magic-lattice spectroscopy was demonstrated on the $^1S_0-{}^3\!P_1\,(M=0)$ transition of $^{88}$Sr \cite{IdoKat03}.
Figure \ref{Fig:LDR}(a) shows the laser induced fluorescence of atoms trapped in the magic lattice, indicating a slight saturation-broadened ($\approx 11$ kHz) spectrum for its natural linewidth of 7.6~kHz.
When the lattice potential was turned off (dataset (b) in Fig.~\ref{Fig:LDR}), the Doppler width  corresponding to the atomic temperature of 6 $\mu$K and the photon recoil shift of
5 kHz appear. In the lattice the recoil was absorbed by the lattice potential and there were no Doppler shifts.
Despite being appealing as a new type of a neutral-atom clock, the serious drawback of this system was its sensitivity to the light polarization of the lattice laser. Indeed, the scalar, vector and tensor contributions to Eq.~(\ref{Eq:alphaBreakDown}) are expectation values of irreducible tensor operators of ranks 0, 1, and 2 respectively.
Due to the angular selection rules, the ground $J=0$ state has only the scalar polarizability, while the excited $J=1$ state
acquires additional vector and tensor contributions. The vector polarizability couples to a residual circular polarization of the lattice and
this substantially increases the clock uncertainty~\cite{IdoKat03}.  The solution was to move to the $^1S_0-{}^3\!P_0$
clock transition, where both states are of a purely scalar nature~\cite{Kat02}.

The lifetime of the $^{3}\!P_{0}$ state determines the natural
width of the clock transition between the ground and the $^3\!P_0$
state. For all \emph{bosonic} isotopes of divalent atoms, the nuclear
spin $I$ vanishes and these isotopes lack hyperfine structure. For
bosonic isotopes the $^3\!P_0$ state may decay only via very weak
multi-photon transitions. However, for the
\emph{fermionic} isotopes, $I \neq 0$,
a new radiative decay channel becomes available due to the hyperfine
interaction (HFI). The HFI admixes $J=1$ atomic states
opening a fast electric-dipole decay route. The resulting HFI-induced decays
determine the lifetimes of the $^3\!P_0$ states and set the natural width
of the clock transition. The HFI-induced rates for fermionic isotopes
were computed by \cite{PorDer04}: a typical value of the radiative width is about 10 mHz.
In bosonic isotopes, lacking the HFI, the transition rate is strongly suppressed as the radiative
decay requires two photons. In this case, the clock transition may be observed by applying
a static magnetic field~\cite{TaiYudOat06}. The B-field admixes the  $J=1$ state of the fine-structure manifold to the $^3P_0$ level, opening the electric-dipole decay channel. In this technique, the magnitude of the clock transition moment may  be experimentally adjusted.

In designing atomic clocks with many atoms, the control and prevention of atomic interactions is of concern.
The collisional frequency shift of atomic clocks operated with ultracold atoms is due to to the mean field energy shift  $\delta E_\mathrm{m.f.}=4 \pi \hbar^2 a n g^{(2)}(0)/M$ of the relevant electronic state, with the
$s$-wave scattering length $a$, atomic number density $n$, and atomic mass $M$.
Here, $g^{(2)}(0)$ is the two-particle correlation function at a zero distance; it is zero for identical fermions and $1\leq g^{(2)}(0)\leq 2$ for distinguishable or bosonic atoms.
Hence, the collisional shifts are suppressed for ultracold fermions, while they are intrinsically unavoidable in bosons.
The quantum statistical nature of atoms is determined by their total spin; that is, bosons have integer spins and fermions have half-integer spins.
In particular, for atoms with an even number of electrons in a $J=0$ state suitable for optical lattice clocks, their nuclear spins $I$ may be zero for bosons and  $I\geq 1/2$ for fermions.
Consequently, the total angular momentum $F=J+I$ of the clock states can be zero for bosonic atoms, but not for fermions, where their coupling to the light polarization of the lattice field is problematic.

Let us  consider two representative lattice geometries for realizing optical lattice clocks.
A one-dimensional (1D) (see Fig.~\ref{Fig:1D}(a)) or 2D lattice composed of a single electric field vector realizes spatially uniform light polarization. In contrast, a 3D lattice  requires at least two electric field vectors; therefore, the synthesized field exhibits a polarization gradient that  varies in space depending on the intensity profile of the lattice lasers.
We discuss that these characteristics of light polarization lead to two optimal lattice clock configurations when  combined with the quantum statistical properties of atoms.

\begin{figure}
\begin{center}
\includegraphics*[width=0.8\linewidth]{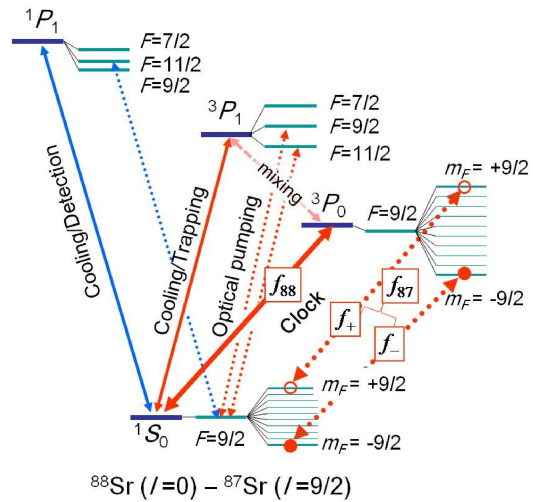}
\end{center}
\caption{(Color online) Energy levels for $^{88}$Sr and $^{87}$Sr atoms. Spin-polarized ultracold $^{87}$Sr atoms were prepared by optical pumping on the $^1\!S_0 (F=9/2) - {}^3\!P_1 (F=9/2)$ transition at $\lambda = 689$~nm with circularly polarized light. The first-order Zeeman shift and the vector light shift on the clock transition at $\lambda = 698$~nm were eliminated by averaging the transition frequencies $f_{\pm}$.}
\label{SrLevel}
\end{figure}

Since the first demonstration, optical lattice clocks were mostly realized with 1D optical lattices employing fermionic~\cite{TakKat03,BruLeTBai06,LudBoyZel06} or bosonic~\cite{BarHoyOat06,BaiFouLeT07} isotopes.
Collision shifts may exist in the 1D scheme with bosonic~\cite{LisWinMid09} or unpolarized fermionic~\cite{CamBoyTho09} atoms because of the relatively high atomic number densities of up to $10^{11}~{\rm cm}^{-3}$ at a single lattice site, which would surely dominate the  uncertainty budget in the future.
Application of spin-polarized fermions~\cite{TakKat09,Gib09} may minimize the collisional frequency shift due to  their quantum-statistical properties.
Figure~\ref{Fig:1D}(a)  shows the schematic diagram for the ``spin-polarized'' 1D optical lattice clock~\cite{TakHonHig06}, where the upward arrows correspond to spin-polarized fermionic atoms.
An advantage of the 1D optical lattice is that the light field polarization is spatially uniform, which allows  canceling out the vector light shift by alternately interrogating the transition frequencies $f_{\pm}$, corresponding to two outer Zeeman components $^1S_0 (F=9/2 , \pm m_F=9/2) - {}^3P_0 (F=9/2, \pm m_F=9/2)$ of the clock transition (see Fig.~\ref{SrLevel})~\cite{TakHonHig06}, to obtain the transition frequency $f_0=\frac{f_{+}+f_{-}}{2}$.
This vector light shift cancellation technique also cancels out the Zeeman shift, thereby realizing virtual spin-zero atoms.

To suppress atomic collisions, the application of  3D optical lattices with less than a single atom in each lattice site, as shown in Fig.~\ref{Fig:LatticeClock}(a), is a straightforward solution.
However, as mentioned earlier, light polarization inhomogeneity inevitable in 3D optical lattices makes a vector light shift for atoms with its angular momentum $F\neq0$  problematic, as the ``vector light shift cancellation'' technique is no longer applicable.
From this viewpoint, the 3D lattice will be  suitable for bosonic atoms with scalar states $(J=0)$.
Technically,  it is more challenging to realize stable 3D optical lattices, as regards their position as well as local polarization, than 1D ones.
One of the simple solutions is to apply ``folded optical lattices''~\cite{RauSchGom98}, where the 3D optical lattice consists of a single standing wave of light. In this configuration, the local polarization of lattice sites remains unchanged, as the two orthogonal electric field vectors oscillate in phase at the local lattice site.
The three-dimensional optical lattice clock has been demonstrated with bosonic $^{88}$Sr atoms \cite{AkaTakKat08,AkaTakKat10}.

\begin{figure}
\begin{center}
\includegraphics*[width=0.8\linewidth]{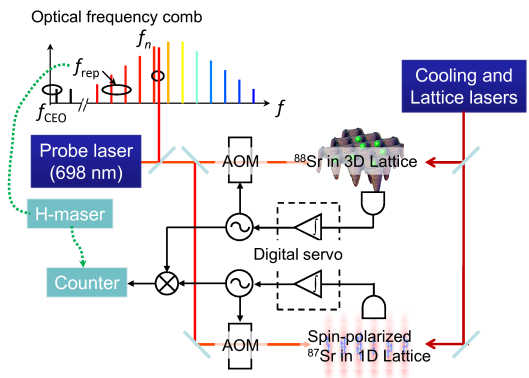}
\end{center}
\caption{(Color online) Two optical lattice clocks with different isotopes and lattice configurations were operated to investigate  their beat note.}
\label{Fig:comp}
\end{figure}

Performance of optical clocks can be  evaluated by comparing two optical clocks with similar performances \cite{ChoHumKoe10,LudZelCam08}, as the state-of-the-art  optical clocks well surpass the primary-frequency standard, Cs clocks, in accuracy as well as stability. Systematic uncertainties of optical lattice clocks were investigated by operating a spin-polarized 1D optical lattice clock with fermionic $^{87}$Sr atoms and a 3D optical lattice clock  with bosonic $^{88}$Sr atoms~\cite{AkaTakKat08}. Figure~\ref{Fig:comp} shows the experimental setup. The clock frequencies of these two atoms differ by the isotope shift of about 62 MHz. Optical lattice clocks with $^{87}$Sr and $^{88}$Sr atoms were alternately prepared and interrogated by clock lasers detuned by the isotope shift. The clock lasers were then servo-locked to the respective transition frequencies of $f_{87}$ and $f_{88}$, and the frequency difference $f_{88} - f_{87}$ was recorded as a time series. The Allan deviation evaluated by the beat note of these two ``independent'' clocks reached $5\times10^{-16}$ for an averaging time of 2,000 s. After a careful elimination of systematic uncertainties, in particular, for the $^{88}$Sr optical lattice clock that was strongly perturbed by mixing magnetic field, the isotope shift on the clock transitions was determined to be $f_{88} - f_{87} = 62,188,138.4(1.3)$ Hz \cite{AkaTakKat08}.

Currently, when the 1D and 3D clocks are compared,  the stability of  optical lattice clock is typically  $10^{-14}/\sqrt{\tau}$ (here $\tau$ is expressed in seconds).
The stability is severely limited by the Dick effect which arises due to the short interrogation time $T_i=$ 60 ms compared to the long cycle time $T_{cyc}=$ 1 s (for a single clock operation) to 6 s (for comparing two clocks sequentially), most of which was spent on cooling and capturing atoms. This situation is equivalent to measuring a laser frequency with a counter having a gate time $T_i$ every $T_{cyc}(>T_i)$. Frequency fluctuations higher than the Nyquist frequency $f_N = 1/(2T_{cyc})$ disturb the measurement by aliasing, as the frequency noise at around the cycle frequency $1/T_{cyc}$ and its harmonics higher than the Nyquist frequency $f_N$ are down-converted into lower frequencies $f\ll1/T_{cyc}$. This aliasing noise, mixed in with the error signal in a feedback loop, causes a long-term white frequency noise in the stabilized laser~\cite{SanAudMak98}. Time-consuming lattice reloading process may be avoided if minimally destructive or quantum non-demolition (QND) schemes \cite{LodWesLem09} are applied to the state detection of the clock transition, as the schemes prevent atoms from being heated out of the lattice trap and allow the reuse of trapped atoms, as in the case of single-ion based clocks.
These improvements will allow us to explore the fractional uncertainties of $10^{-17}$ in a reasonable averaging time of a few 100 s.

At a few $10^{-17}$ fractional uncertainties,  shifts due to black-body radiation (BBR) \cite{PorDer06}  will dominate the uncertainty budget of Sr-based lattice clocks.
As the BBR shift rapidly decreases as $T^{4}$ for a surrounding temperature $T$, a cryogenic environment, even at the liquid nitrogen temperature of $T$ = 77 K, will reduce the BBR shifts to 10 mHz \cite{KatTakPal03}. Therefore, the corresponding uncertainty will have an effect only at the $10^{-18}$ level or below.
Such a cryogenic environment may be readily applicable to optical lattice clocks, as a cryogenic region, a few cubic millimeters in volume, is sufficient for their operation.
For transferring atoms into the tiny cryogenic volume, a moving magic optical lattice \cite{KisHacFuj06} may  be employed. Such an experiment is now in progress in Tokyo.

Most of the discussed uncertainties  such as the collisional shifts, the BBR shifts, and hyperpolarizability effects depend on  atomic parameters; therefore, they can be improved by a proper choice of specific atom.
The optical lattice clock scheme is generally applicable to atoms of groups II and IIb \cite{OvsPalKat06} such as Ca~\cite{DegStoRie04}, Yb~\cite{PorDerFor04}, Zn, Cd~\cite{YeWan08}, and Hg~\cite{HacMiyPor08} that have  hyperfine-mixed $J=0\rightarrow J=0$ transition between long-lived states.
Alternatively, a multiphoton excitation of the clock transition \cite{SanAriIdo05,HonCarNag05}, or the mixing of the $^3P_0$ state with the $^3P_1$ state using a magnetic field \cite{TaiYudOat06} or an elliptically polarized light~\cite{OvsPalTai07} may allow the use of even isotopes that exhibit purely scalar nature of the $J=0$ state.
The optical lattice clock with the best performance needs to be experimentally explored among possible candidates because of difficulties in predicting some of the uncertainties associated with higher-order light field perturbations, such as resonant contributions to the 4th-order light shifts and multiphoton ionization processes.

\section{MICROMAGIC CLOCKS}
\label{Sec:MuMagic}
The optical lattice clocks described in the previous section operate at optical frequencies,
requiring frequency combs to convert the optical frequencies to the microwave domain suitable for counting time.
Meanwhile, for the past four decades, the
second has been defined in terms of the microwave transition in $^{133}$Cs atom.
Cs clocks serve as primary frequency standards worldwide
and there is a substantial investment in the infrastructure supporting these clocks.
The most accurate Cs clocks are fridge-sized fountain clocks (see, e.g., Ref.~\cite{WynWey05}).
The length of a meter-long active chamber is determined by requiring that the clock interrogation time (i.e., the time it takes the  atoms to fly up and down the chamber in the gravitational field) does not limit the spectroscopic resolution.
By contrast, developing {\em microwave} lattice clocks  may be beneficial as the active chamber
of the clock will be reduced to a few micrometers across. This million-fold reduction in size
is anticipated to lead to a better control over detrimental black body radiation and stray magnetic fields.  In addition, the hyperfine manifolds are used to
store quantum information in a large fraction of quantum computing
proposals with ultracold alkalis. Finding magic conditions would enable decoherence-free trapping for these important realizations of qubits.

As discussed in the introduction, presently the stability of the primary Cs frequency standard is limited by the quantum projection noise (QPN) limit~\cite{ItaBerBol93}, described by Eq.~(\ref{Eq:Allan})~\cite{SanLauLem99}.
The stability can be substantially improved by using techniques from quantum information processing.
It may be shown (see e.g., \cite{LeiBarSch04}) that the stability of highly-entangled ensemble of $N_\mathrm{at}$ atoms scales as $1/N_\mathrm{at}$ versus the
QPN scaling~(\ref{Eq:Allan}) of $1/\sqrt{N_\mathrm{at}}$. For a sample of a million atoms the measurement time would be reduced by a factor of a thousand.  Over the past decade, Cs and Rb atoms were studied as candidates for quantum computing in optical traps (see~\cite{SafWalMol10} and references therein)
and the developed quantum logic is applicable to entangling ``MicroMagic'' clocks. The lattice clocks may harness the power of entanglement for improving  the stability of microwave clocks.

The idea of microwave lattice clocks was discussed by \citet{ZhoCheChe05} for $^{133}$Cs. However, these authors have misidentified magic trapping conditions; a more sophisticated theoretical analysis and an experimental study~\cite{RosGheDzu09}
have rendered conclusions of that paper invalid. Based on the detailed understanding of atomic AC polarizabilities,
\citet{BelDerDzu08Clock} found that the magic trapping conditions, however, can be attained for aluminum or gallium atoms.
These authors have coined a term ``microMagic'' clock to emphasize both the micrometer size of the trap and
the microwave frequency of such a clock.
Further work on Cs revealed that the magic conditions may be ultimately attained by introducing additional
``magic'' magnetic fields and ``magic'' angles between magnetic fields and the axis of a 1D optical
lattice~\cite{FlaDzuDer08,LunSchPor10,Der10Bmagic,Der10DoublyMagic}.
Below we review these developments.

The microwave clockwork involves two atomic levels of the same
hyperfine manifold attached to an electronic state $nJ$.
Hyperfine splittings primarily arise due to an interaction
of atomic electrons with the nuclear magnetic moment.
We consider atoms trapped in a 1D optical
lattice formed by either linearly of circularly-polarized lasers. The quantizing magnetic
field $\mathbf B$ in general may be directed at some angle to the lattice.
The clock states are commonly labeled as $\left\vert
n\left( IJ\right) FM_{F}\right\rangle $, where $I$ is the nuclear
spin, $J$ is the electronic angular momentum, and$\ F$ is the total
angular momentum, $\mathbf{F}=\mathbf{J}+\mathbf{I}$, with $M_{F}$
being its projection on the quantization axis. For simplicity, below we will focus on the $J=1/2$ electronic
states. Then for the $I \neq 0$ isotopes there are two hyperfine structure
states $F'= I+1/2$ and $F=I-1/2$.
We will use the short-hand notation $|F\rangle$ and $|F'\rangle$ for the lower and  upper clock states.   In particular, for the $^{133}$Cs atom ($I=7/2$) the clock transition is
between the $F=4$ and $F=3$ hyperfine components of the  $6s_{1/2}$ electronic
ground state.

We would like to find a magic wavelength for a hyperfine transition by requiring that
$\alpha_F'(\omega_m) - \alpha_F(\omega_m)=0$. At this point one may evaluate the dynamic polarizabilities and deduce the magic wavelength. However, such calculations require additional care. Indeed, we are considering the Stark shift of hyperfine levels attached to the same electronic
state. To the leading order, the shift is determined by the properties of the underlying electronic state. However, because the electronic state for
both hyperfine levels is the same, the scalar Stark shift of both levels is the same.
An apparent difference between the two
clock levels is caused by the hyperfine interaction (HFI), and the rigorous analysis involves
so-called HFI-mediated polarizabilities~\cite{RosGheDzu09}.

Qualitatively, the importance of a consistent treatment of the  HFI-mediated polarizabilities may be understood by considering the expression for the scalar polarizability,
\begin{equation*}
\alpha_{nF}^S\left( \omega\right) =\frac{1}{3}\sum_{i} \sum_{p=x,y,z}\frac{\langle nFM_{F}|D_{p}|i\rangle \langle i|D_{p}|nFM_{F}\rangle }{E_{nFM_{F}}-E_{i}+\omega} +...
\end{equation*}
where the omitted term differs by $\omega \rightarrow -\omega$, and $D_p$ is a component of the dipole operator. All the involved states are the hyperfine
states. While this requires that  the energies include hyperfine splittings, it also  means
that the wave-functions incorporate HFI to all-orders of perturbation theory.
Including the experimentally-known hyperfine splittings in the
summations is straightforward and some practitioners (see, e.g., \cite{ZhoCheChe05})  may stop at that, completely neglecting the HFI corrections to the wave-functions. This is hardly justified as
both contributions are of the same order. Lengthy third-order (two dipole couplings
to the laser field and one HFI) expressions for these polarizabilities may be found in Ref.~\cite{RosGheDzu09}. In the following, we  keep the symbol $\alpha$ for the traditional second-order polarizabilities and use $\beta$ for the HFI-mediated polarizabilities.

The clock transitions in divalent atoms are between non-magnetic states; this removes sensitivity to magnetic fields.
For $J\neq 0$ atoms like Cs, however, there is  an additional piece of the puzzle: the clock states are sensitive to both optical and magnetic fields.
One needs to eliminate the sensitivity of transition frequency $\nu$ to both perturbations simultaneously. Below we
 consider two possibilities to remove the sensitivity to the Zeeman effect: (i) work with $M_F'=M_F=0$ magnetic substates in very weak B fields, thereby eliminating the Zeeman shift and (ii) operate on $M_F \rightarrow  - M_{F}$ transitions; such transitions have ``magic'' values of $B$ fields, where the Zeeman sensitivity is removed.

\subsection{ $M_F=0 \rightarrow M_{F'}=0$ clock transitions}
For the $M_F=0$ hyperfine sublevels and linear polarization of the lattice laser, the
vector contribution to polarizability ~(\ref{Eq:alphaBreakDown}) vanishes.
We find that the differential polarizability may be parameterized as
\begin{equation}
\Delta \alpha \left(  \omega_{L}\right)  =   A\left(  F^{\prime}%
,F\right)  \beta^S_{ F}\left(  \omega_{L}\right)
+  B\left(  F^{\prime},F\right)  \beta^T_{F}\left(  \omega_{L}\right)
  , \label{Eq:ClockShiftSandT}%
\end{equation}
where pre-factors $A$ and $B$ depend on the $F$-numbers of
the clock states and on the orientation of the
quantizing B field. The relation
(\ref{Eq:ClockShiftSandT}) arises due to the fact that the
respective scalar and tensor parts of the dynamic polarizability
vary proportionally for the two clock states. Clearly the scalar and
tensor contributions to the differential shift must cancel each
other at the magic wavelength.

We start with discussing the results for the metrologically
important $^{133}$Cs atom. Calculations and experiment~\cite{RosGheDzu09} find
that there is no magic wavelength for the Cs clock.
A partial solution to this problem was found~\cite{LunSchPor10} (see \cite{Der10Bmagic} for theory):
one needs to apply a relatively large (a few Gauss) bias magnetic field of a specific value making trapping ``magic'' for a given trapping laser wavelength. As a result, however, the transitions becomes Zeeman-sensitive
through the second-order effects; numerical estimates show that, unfortunately, the residual B-field sensitivity would preclude designing a competitive clock.

Qualitatively, for Cs, the tensor contribution to the clock shift is
much smaller than the scalar contribution and this leads to
unfavorable conditions for reaching the cancelation of the scalar
and tensor shifts in Eq.~(\ref{Eq:ClockShiftSandT}).
To cancel the light shift we need to
find atoms where the scalar and tensor shifts are comparable. This
happens for  atoms having the valence electrons in the $p_{1/2}$
state. For non-zero nuclear spin, the $p_{1/2}$ state has two
hyperfine components that may serve as the clock states.  The advantage of
the $p_{1/2}$ state comes from the fact it is part of a
fine-structure manifold: there is a nearby $p_{3/2}$ state separated
by a relatively small energy interval determined by the relativistic
corrections to the atomic structure. This small interval amplifies the
tensor part of the polarizability and does not affect the scalar contribution
to Eq.~(\ref{Eq:ClockShiftSandT}).

Based on these qualitative considerations,
\citet{BelDerDzu08Clock} found magic wavelengths for Al and Ga atoms.
These are group III atoms with the $p_{1/2}$ ground state. For example, in  $^{27}$Al the clock
transition is between the hyperfine structure levels $F=3$ and $F=2$
in the ground $3p_{1/2}$ state. The
clock frequency is about 1.5 GHz, placing it in the microwave region. Furthermore, cooling Al has already been demonstrated \cite{McGGilLee95}.
\begin{figure}[h]
\begin{center}
\includegraphics*[width=0.8\linewidth]{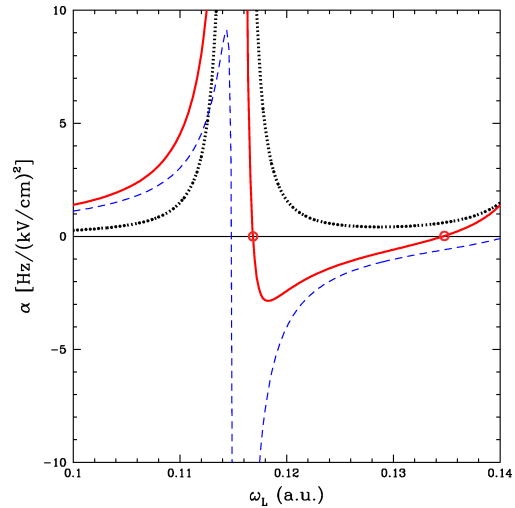}
\end{center}
\caption{(Color online) Differential polarizability for Al
$\mu$Magic clock in the $\mathbf B \parallel \hat k$  geometry as a
function of the lattice laser frequency. Dotted line: contribution
from the scalar term; dashed line: contribution from the tensor term;
solid line: total polarizability. Total clock shift vanishes at two
``magic'' values of the laser frequency. } \label{Fig:AlCancellation}
\end{figure}
The cancelations between scalar and tensor
contributions to the clock shift in Eq.~(\ref{Eq:ClockShiftSandT})
is illustrated in Fig.~\ref{Fig:AlCancellation}. There are two magic wavelength for the geometry $\mathbf B \parallel \hat k$. \citet{BelDerDzu08Clock} have carried out estimates for various factors affecting
such a clock similar to discussion of Sec.~\ref{Sec:Optical}. They concluded that the proposed microwave lattice (microMagic) clock may  compete with the state of the art
fountain clocks.

%

\subsection{ $M_F \rightarrow  - M_{F}$ clock transitions}

\begin{figure}[h]
\begin{center}
\includegraphics*[width=0.8\linewidth]{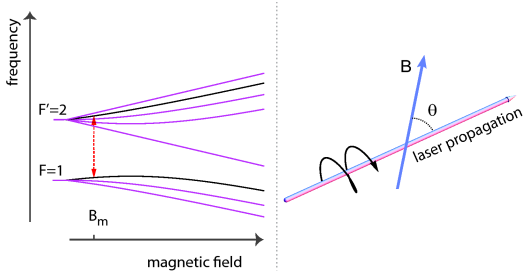}
\end{center}
\caption{(Color online)
Left panel: Zeeman effect (Breit-Rabi diagram) for the hyperfine manifold in the ground state of $I=3/2$ isotopes of alkalis. Two clock levels $|F'=2, M'_F=+1 \rangle$ and  $|F=1,M_F=-1\rangle$ are shown in
black. Clock transition at the magic B-field is indicated by a vertical red doubly-headed arrow.
Right panel illustrates geometry of laser-atom interaction: degree of circular polarization, angle $\theta$, and laser wavelength may be varied.
 \label{Fig:GeometryZeeman}
 }
\end{figure}

To reiterate, for weak B fields there are no magic conditions on the
$M_F=0 \rightarrow M_{F'}=0$ clock transitions in $^{133}$Cs. As we move to magnetic substates,
we require that at the magic B-field $d\nu/dB (B_m) = 0$.
Such conditions occur, for example, for a two-photon
$|F'=2, M'_F=+1 \rangle \rightarrow |F=1,M_F=-1\rangle$ transition  in $^{87}$Rb
at the field of about 3 Gauss.
The relevant Breit-Rabi diagram is shown in Fig.~\ref{Fig:GeometryZeeman}.
The two clock levels are highlighted: the existence of the magic B-field may be inferred visually.
The existence of the Stark-Zeeman  ``doubly-magic'' conditions was found by \citet{Der10DoublyMagic}; the discussion below is based on that paper.

For the $M_F \rightarrow  - M_{F'}$ transitions,
the electronic $g$-factors of the
two states are the same (see Fig.~\ref{Fig:GeometryZeeman}). Then the bulk of the Zeeman shift of the transition frequency goes away and the linear Zeeman effect is determined only by the nuclear $g$-factor $g_I = 1/I \,\mu_{\rm nuc}/\mu_N$, where $\mu_N$ is the nuclear magneton. This residual linear shift is comparable to the second-order (in electronic magnetic moment) Zeeman correction, quadratic in the B-field. By evaluating
the derivative of the total (linear + quadratic) shift with respect to magnetic field we find that the Zeeman shift goes through a minimum at the following ``magic'' value of the B-field
\begin{equation}
 B_{m}\approx
\frac{g_{I}\mu _{N}~M_{F^{\prime }}}{2\left\vert \langle F,M_{F^{\prime }}\left\vert \mu _{z}^{e}\right\vert F^{\prime },M_{F^{\prime }}\rangle \right\vert ^{2}} \, h\nu_0 \,.
\label{Eq:Bmagic}
\end{equation}
Here $\mu^e$ is the electron magnetic moment operator.
Values of $B_m$ for Rb and Cs isotopes are tabulated in Table~\ref{Tab:BmList}.
These fields are relatively weak and can be well
stabilized using existing technologies~\cite{LacReiRam10}.

\begin{table}[h]
\caption{ Values of ``magic'' B-fields and ranges of ``magic'' wavelengths for metrologically important
$^{133}$Cs and $^{87}$Rb.
 \label{Tab:BmList}}
\begin{ruledtabular}
\begin{tabular}{ccc}
Transition  &
$B_m$, Gauss & $\lambda_m$ \\
\hline
\multicolumn{3}{c}{
$^{87}$Rb,  $I=3/2$,  $\nu_0=6.83 \, \mathrm{GHz}$} \\
$|2,1\rangle \to  |1,-1\rangle$ & 3.25  & 806 nm\footnotemark[1]  \\[1ex]
\multicolumn{3}{c}{
$^{133}$Cs,  $I=7/2$,  $\nu_0=9.19 \, \mathrm{GHz}$} \\
$|4,1\rangle \to  |3,-1\rangle$ & 1.41   &  --- \\
$|4,2\rangle \to  |3,-2\rangle$ & 3.51  & 906--1067; 560-677\\
$|4,3\rangle \to  |3,-3\rangle$ & 9.04  & 898--1591;863--880; 512--796\\
\end{tabular}
\end{ruledtabular}
\footnotetext[1]{nearly doubly-magic}
\end{table}

Fixing magnetic field at its magic value accomplishes the Zeeman-insensitivity of the clock transitions.
Now we would like to additionally remove the Stark sensitivity to  intensity of trapping
laser fields.  We consider the following setup shown in Fig.~\ref{Fig:GeometryZeeman}.
An atom is illuminated by a circularly-polarized laser light.
At the same time, a bias magnetic field is applied at an angle $\theta$ to the direction
of laser propagation. The B-field is fixed at its magic value. This is a basic building block for optical trapping.

The differential polarizability in this case reads
\begin{eqnarray}
\lefteqn{\Delta \alpha  (\omega)=
\left(  \beta_{F^{\prime}}^{S}-\beta_{F}^{S}\right)  +}
\label{Eq:ClockShiftMultiPhoton}
\\
&&  \mathcal{A}\cos\theta~M_{F^{\prime}} \left[  \left(  \frac
{1}{2F^{\prime}}\beta_{F^{\prime}}^{V}+\frac{1}{2F}\beta_{F}^{V}\right)
+
g_{I}\frac{\mu_{N}}{\mu_{B}}~\alpha_{nS_{1/2}}^{V} \right]  \, . \nonumber
\end{eqnarray}
The last contribution arises due to an interference between Stark and Zeeman interactions. Qualitatively, the vector contribution to the Stark shift has the very same rotational
properties as the Zeeman coupling (both are vector operators). These operators, in particular,
couple the two hyperfine manifolds. Consider the shift of the $|F',M_F\rangle$ level. The Zeeman operator couples it to the $|F,M_F\rangle$ intermediate state, and then the vector Stark shift operator
brings it back to the $|F',M_F\rangle$ level thereby resulting in the energy shift. This cross-term is of the same order of magnitude
as the other two terms in Eq.~(\ref{Eq:ClockShiftMultiPhoton}) and has to be included in the consideration.

Since the magic condition corresponds to $\Delta \alpha (\omega_m) =0$, we may recast
Eq.~(\ref{Eq:ClockShiftMultiPhoton}) into
\begin{equation}
M_{F^{\prime }} \mathcal{A}\cos \theta =-\frac{\beta _{F^{\prime }}^{S}-\beta _{F}^{S}}{\left( \frac{1}{2F^{\prime }}\beta _{F^{\prime }}^{V}+\frac{1}{2F}\beta _{F}^{V}\right) +
g_{I}\frac{\mu _{N}}{\mu _{B}}\alpha _{nS_{1/2}}^{V}} \, .
\label{Eq:MFAcos}
\end{equation}
The r.h.s.\ of this equation depends on the laser frequency, while the l.h.s.\ does not.
Moreover, $|\mathcal{A}\cos \theta| \le 1$, therefore the magic conditions would exist
only if for a given $\omega$  the r.h.s.\ is within the range $-|M_F'|$ and $|M_F'|$.

\begin{figure}[h]
\begin{center}
\includegraphics*[width=0.8\linewidth]{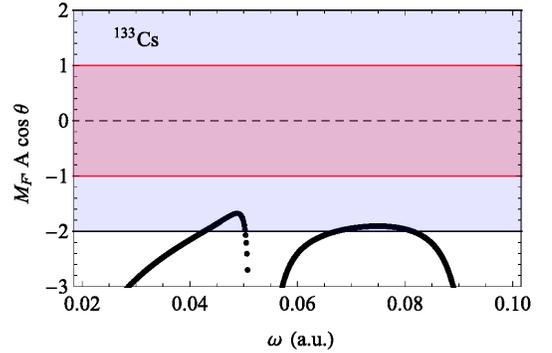}
\end{center}
\caption{(Color online) Magic conditions for $^{133}$Cs. A dependence of the product
$M_{F^{\prime }}A\cos \theta$ on trapping laser frequency (in atomic units) is plotted. The shaded regions
are bound by $-|M_{F^{\prime }}|$ and $+|M_{F^{\prime }}|$ lines. Magic trapping for
a $|F'=4,M_F'\rangle \to |F=3,-M_F'\rangle$ clock transition is only possible when
the computed curve lies inside the corresponding shaded region.
 \label{Fig:133CsAcos}
 }
\end{figure}

From Fig.~\ref{Fig:133CsAcos} we see that the doubly-magic trapping of $^{133}$Cs  atoms
is indeed possible for two transitions: $|4,2\rangle \to  |3,-2\rangle$ and
$|4,3\rangle \to  |3,-3\rangle$. The only complication is that driving the former transition requires 4 photons,
while the latter transition requires 6 photons. This may be potentially
accomplished either with multi-step RF or MW or stimulated Raman drives~\cite{HarLewMcG02,AlePaz97}.
Similar analysis for $^{87}$Rb $|F'=2,1\rangle \to |F=1,-1\rangle$ transition shows that the  $M_{F^{\prime }}\mathcal{A}\cos \theta$ curve nearly touches its limiting value at
$\lambda_m \approx 806\, \mathrm{nm}$. Here the r.h.s.\ of Eq.~(\ref{Eq:MFAcos}) reaches values of
$\approx -1.05$, i.e., it is just 5\% off the limiting value of -1. While not
quite achieving the ``doubly-magic'' status, this 806 nm wavelength gets us to nearly-magic conditions. Recently this prediction was verified experimentally~\cite{ChiNelOlm10}.

%

\section{BEYOND TIME-KEEPING}
\label{Sec:Beyond}

In this concluding section, we will present several examples of how the exquisite accuracy and stability of the lattice clocks may be used  for precision measurements and quantum information processing.

Usually the environmental effects (e.g., stray fields) degrade the performance of the clocks. One may turn this around and by measuring shifts of the clock frequency, characterize an interaction with the environment. The most fundamental experiments of this kind search for a potential variation of fundamental constants~\cite{ForAshBer07,BlaLudCam08}, where the ``environmental agent'' is the fabric of the Universe itself, affecting the rate of ticking of atomic clocks. In other experiments one may probe ultracold collision physics~\cite{LudZelCam08,CamBoyTho09} or
map out atom-wall interaction and search for non-Newtonian gravity~\cite{WolLemLam07,SorAlbFer09,DerObrDzu09} by monitoring the clock frequency.

In addition, atomic clock states may serve as a perfect quantum memory (qubit). Good clock states
make also good qubit states, as they are well isolated from detrimental environmental decoherences.
There are several recent proposals~\cite{HayJulDeu07,GorReyDal09,DalBoyYe08} that use optical lattice clocks as a platform for quantum computation and simulation. It is worth noting that the initial developments in quantum information processing (QIP) with atoms dealt with qubit states stored in the hyperfine structure of alkalis. Ideas on magic trapping conditions for microMagic clocks are starting to make an impact~\cite{DudZhaKen10} in QIP, as experimentalists start improving
coherence times using magic trapping techniques.

\subsection{ Time and space variation of fundamental constants}
Some cosmological models and unification theories imply that the fundamental physical
constants (such as the fine-structure constant, $\alpha=e^2/(\hbar c) \approx 1/137$)
may vary with time~\cite{Uza03}. The values of the constants may also depend on local hypothetical couplings
to ambient gravitational or other  fields. These propositions may be probed with atomic clocks.
Indeed, atomic clocks can monitor frequencies of atomic transitions with unprecedented accuracy. The frequencies of two distinct atomic transitions (e.g., microwave and optical) depend differently on fundamental constants.  By comparing outputs of two clocks as a function of
time or position in space, one may deduce limits on space-time variation of fundamental constants.
Spatial dependence, in particular, could be tested as the Earth's elliptic orbit takes the clocks through a varying solar gravitational potential or in satellite-based mission. Although still 	
nascent, the optical lattice clocks have already made an important
contribution to constraining space-time variations.
\citet{BlaLudCam08} have analyzed a three-year record of the $^1\!S_0-\,^3\!P_0$ clock transition frequency in neutral $^{87}$Sr taken
by three independent laboratories in Boulder, Paris, and Tokyo. They combined periodic
variations in the clock frequency with $^{199}$Hg$^+$ and H-maser data and
obtained the strongest limits to date on gravitational-coupling coefficients for the fine-structure constant, electron-proton mass ratio, and light quark mass. In addition,
in combination with the previous atomic-clock data  they have increased confidence
in the zero drift result for the modern epoch.

Optical  clocks are particularly sensitive to variation of the fine-structure constant.
Thus far, 
the most stringent test was carried out by comparing Al$^+$ and Hg$^+$ clocks over two years with a fractional uncertainty of $5 \times 10^{-17}$ to verify the constancy of
$|\dot{\alpha}/\alpha|$ at the level of $(-1.6\pm2.3 )\times 10^{-17}/{\rm yr}$ \cite{RosHumSch08}.
Assuming $\Delta\alpha/\alpha=10^{-16}$ per year, at which level astrophysical determinations have given controversial results\cite{Uza03}, the fractional change in the clock frequency \cite{AngDzuFla04} $\delta \nu/\nu_0$ can be $6.2\times 10^{-18}$, $3.1\times 10^{-17}$, and $8.1\times 10^{-17}$ for Sr-, Yb-, and Hg-based optical lattice clocks, respectively.
Heavier atoms such as Yb and Hg are more sensitive to $\alpha$ as the underlying relativistic corrections scale as the  nuclear charge squared. One may envision taking a Sr lattice clock as an anchor and detecting the fractional frequency change of the Hg lattice clock at the $\delta \nu/\nu_0 = 10^{-17}$ level, which can be accurately measured with an optical frequency comb technique.

\subsection{Atom-wall interaction}
\begin{figure}[h]
\begin{center}
\includegraphics*[width=0.8\linewidth]{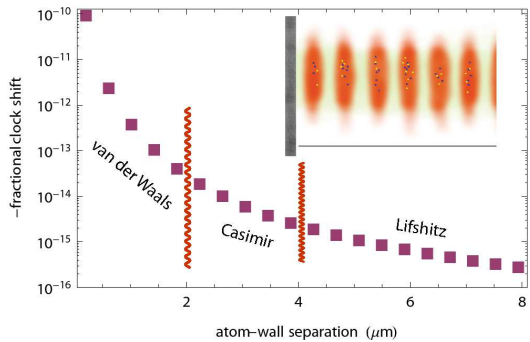}
\caption{(Color online)  Fractional clock shifts for Sr as a function of separation from a gold surface at  $T=300 \,\mathrm{K}$.   Individual points represent shifts in individual trapping sites of the optical lattice. First well is placed at $\lambda_m/4 \approx 200 \, \mathrm{nm}$ and subsequent points are separated by $\lambda_m/2 \approx 400 \, \mathrm{nm}$. Inset: Idealized setup for measuring atom-wall interaction with optical lattice clocks.
\label{Fig:atom-wall}}
\end{center}
\end{figure}

An idealized setup for measuring atom-wall interaction with lattice clocks is shown in Fig.~\ref{Fig:atom-wall}.  A conducting surface of interest acts as a mirror for the laser beam normally incident on the surface. The resulting interference of the beams forms an optical lattice. Laser operates at a ``magic'' wavelength $\lambda_m$. One could work with a 1D optical lattice for which the atoms are attracted to the laser intensity maxima. The first pancake-shaped atomic cloud would form at a distance $\lambda_m/4$  from the mirror. The subsequent adjacent clouds are separated by a distance $\lambda_m/2$.  By monitoring the clock shift at individual trapping sites, one measures a distance dependence of the atom-wall interaction.

As the separation $z$ between an atom and a wall increases, the atom-wall
interaction evolves  through several distinct regimes: (i) chemical-bond
region  that extends a few nm from the surface, (ii) van der Walls region ($V \propto z^{-3}$),
(iii) retardation (Casimir-Polder) region ($V \propto z^{-4}$), and (iv) Lifshitz (thermal bath fluctuations) zone ($V \propto z^{-3}$). Due to the interaction with the wall, the clock levels would shift.
The computed fractional clock shift for Sr clock \cite{DerObrDzu09} is shown in Fig.~\ref{Fig:atom-wall}. The spatially-separated zones of the the three regimes of the long-range interaction are shown.
We immediately see that the atom-wall interaction is a large effect, corresponding to $10^{-10}$ fractional clock shifts at the first well. This is roughly a millon time larger than the demonstrated accuracy of the Sr clock~\cite{LudZelCam08}. Moreover, lattice clocks can be used to detect all three qualitatively-distinct mechanisms of the atom-wall interaction.
In this regard, the lattice clocks offer a unique
opportunity to map out both van der Walls$\rightarrow$Casimir-Polder and Casimir$\rightarrow$Polder-Lifshitz transition regions.
This distinguishes the lattice clock proposal from
previous experiments: the former transition was probed by \citet{SukBosCho93}, while the latter was detected by \citet{ObrWilAnt07}.
None of the experiments so far has been able to map out both transitions simultaneously.

\subsection{Entangling the lattice clock}
A number of proposals have noted the virtues of using alkaline-earth-like atoms in lattices for quantum information and quantum computing \cite{DerCan04,HayJulDeu07,GorReyDal09,DalBoyYe08,ShiKatYam09}. Below we highlight using quantum-information concepts such as entanglement for improving the atomic clock.
Quantum entanglement is a crucial resource in quantum computing
and has the potential to improve precision
measurements~\cite{NieChu00,ChiPreRen00}.
\citet{WeiBelDer10} proposed a scheme for entangling an optical lattice clock,
with the specific goal of demonstrating the power of entanglement for measuring time.

Measuring time  with atoms relies on the fact that the quantum-mechanical probability
of making a transition between two clock levels depends on the detuning $\Delta \nu$
of the  probe field $\nu$ from the atomic transition frequency $\nu_0$.
By measuring the probability as a function of $\nu$, one can infer if the two frequencies
are equal and thereby ``lock'' a local oscillator to the atomic transition.
The precision of measuring  $\Delta \nu$ is limited by the quantum projection noise \cite{ItaBerBol93}.  For a measurement of $N_\mathrm{at}$ unentangled atoms the resulting signal-to-noise of $\Delta \nu$ scales as $\sqrt{N_\mathrm{at}}$:  the standard quantum limit. The  use of entanglement holds the promise of improving clock precision to the Heisenberg limit, with  signal-to-noise scaling as $N_\mathrm{at}$.

\begin{figure}[h]
\begin{center}
\includegraphics*[width=0.8\linewidth]{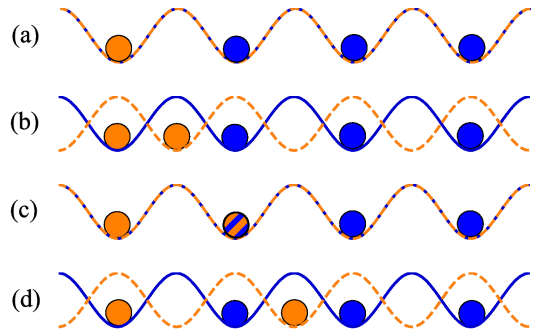}
\end{center}
\caption{(Color online) Schematic of the entanglement process.
The transport lattice is created by a superposition of two displaced circularly polarized standing wave lattices: $\sigma_+$ and  $\sigma_-$ lattices.
(a) A single head atom (orange circle) and several clock atoms
(blue circles) are trapped in the minima of a 1-D optical lattice, with one or fewer atoms per site.
Due to an intensity differential of the underlying lattices, the clock atoms
couple strongly to the $\sigma_+$ lattice (solid blue line). The head
atom is placed in a superposition of atomic states: one which couples strongly to the $\sigma_+$ lattice and one which couples strongly to the $\sigma_-$ lattice (dashed
orange line). (b) As the displacement between  the two circularly-polarized lattices increases, the $\sigma_-$ state is spatially separated and is transported along the lattice. (c)
This portion of the head atom is then brought into contact with a clock
atom to entangle the two atoms. (d) The head atom is transported
further to obtain entanglement with the remaining clock atoms in a
similar manner.
} \label{Fig:transport}
\end{figure}

In the scheme of \citet{WeiBelDer10} the divalent clock atoms
are held in a lattice at a ``magic'' wavelength that does not perturb the clock frequency -- to maintain clock accuracy -- while an open-shell $J=1/2$ ``head'' atom is coherently transported between lattice sites via the lattice polarization.   This polarization-dependent ``Archimedes' screw'' transport at magic wavelength takes advantage of the vanishing vector polarizability of the scalar, $J=0$, clock states of bosonic isotopes of divalent atoms (see Fig.~\ref{Fig:transport}).  The on-site interactions between the clock atoms and the head atom are used to engineer entanglement and for clock readout.
Estimates show that roughly a 1,000 clock atoms can be entangled with this scheme.

Notice that  many of the usual requirements for producing highly-entangled states between atoms -- such as single-site addressability, single-site  readout, and unity site occupation -- are absent in this scheme.
The proposed scheme occupies an interesting ``middle-ground'' of experimental schemes for clock entanglement.  It holds promise for use with larger numbers of atoms than has been demonstrated to date with ion traps\cite{LeiBarSch04}.  And while it cannot entangle as large-number samples as are used in spin-squeezing experiments\cite{KuzBigMan98,SchLerVul09,MeiYeHol08}, it may be able to produce greater levels of entanglement.

To conclude, over a time-span of just a few years since their inception, optical lattice clocks became one of the most accurate
timekeeping devices ever built. Presently, their accuracy and stability surpass the primary frequency standard: optical lattice clocks are contenders for a future redefinition of the second. While  most of the experimental work so far focused on
Sr and Yb atoms, we think that Hg is a promising candidate for a highly-accurate optical lattice clock. The projected fractional accuracy of the Hg clock is $3\times 10^{-19}$~\cite{HacMiyPor08}. This is a few orders of magnitude better than
the accuracy of the present clocks. We have highlighted several applications, e.g., tracking time-space variations of fundamental constants, which may benefit from such accuracies. Still an open question remains what applications, both fundamental and practical, may take advantage of the superb precision and stability of optical lattice clocks.

The fruitful ideas of optical lattice clocks may be extended to the microwave
domain. The work on the microMagic lattice clocks so far has been of a conceptual nature
and the experimental feasibility of such clocks is yet to be studied.
Yet, it is anticipated that a variety of applications could benefit from the magic (and nearly-magic) conditions. For example, we anticipate that lifetimes of quantum memory~\cite{ZhaDudJen09} may be improved. Another interesting opportunity is to co-trap divalent and alkali-metal or group III atoms in the same lattice. For microwave transitions there is usually a range of ``magic'' trapping conditions and
magic wavelengths for divalent atoms may fall within this range. Then one could design a dual-species microwave-optical clock sharing the same lattice \cite{MorDzuDer10}.

\section*{Acknowledgments}

We thank Muir Morrison for comments on the manuscript.
The work of A.D. was supported in part by the U.S. NSF and by U.S. NASA under Grant/Cooperative
Agreement No. NNX07AT65A issued by the Nevada NASA EPSCoR program.
H.K. was supported in part by the Photon
Frontier Network Program of MEXT, Japan, and by the JSPS through its FIRST program.



\end{document}